\documentclass[aip, reprint]{revtex4-2} 

\usepackage{amsmath}  
\usepackage{amsfonts} 
\usepackage{graphicx} 

\usepackage{url}
\usepackage[hidelinks]{hyperref}
\usepackage{bookmark}

\begin{document}


\title{Investigating nonlinear dynamics in a mass-spring oscillator using real-time computer vision}


\author{Marco P. M. de Souza}
\email{marcopolo@unir.br}
\affiliation{Grupo de Pesquisa Estrutura da Matéria e Física Computacional, Universidade Federal de Rondônia, Campus Ji-Paraná, Rondônia, Brazil}


\date{\today}

\begin{abstract}

We present a low-cost laboratory for investigating the dynamics of a vertical mass--spring oscillator using real-time computer vision. The proposed system automatically tracks the motion of the oscillating mass with a smartphone camera, displays the displacement in real time, and exports the recorded data for further analysis. Representative experiments include damped oscillations, oscillation period measurements, spectral analysis, and phase-space reconstruction. The system also enables the investigation of nonlinear phenomena, including harmonic generation, frequency mixing, and energy exchange between coupled oscillation modes. Owing to its low cost, ease of construction, and automated data acquisition, the proposed apparatus extends the traditional mass--spring experiment to include topics in nonlinear dynamics that are rarely explored in undergraduate laboratories.

\end{abstract}

\maketitle 

\section{Introduction} 

The vertical mass-spring oscillator is one of the most widely used experimental systems in introductory physics laboratories. Owing to its conceptual simplicity and inexpensive construction, it provides an accessible framework for introducing fundamental topics such as Hooke's law, simple harmonic motion, and damping. Measurements of the oscillation period provide a straightforward way to investigate the dependence of the period on the oscillating mass and, more generally, to characterize the dynamical behavior of the system.

Despite its apparent simplicity, however, a real mass-spring system exhibits a much richer dynamics than is usually explored in undergraduate laboratory courses. The finite mass of the spring, departures from Hooke's law at larger elongations, and the coexistence of vertical and pendular oscillation modes give rise to nonlinear effects that include higher harmonics, energy transfer between oscillation modes, and frequency mixing. These phenomena provide valuable opportunities to introduce students to concepts that extend beyond the ideal harmonic oscillator while remaining experimentally accessible.

Numerous experimental approaches have been proposed to investigate the dynamics of mass-spring oscillators in undergraduate laboratories. In addition to traditional stopwatch-based measurements, recent developments have employed force sensor \cite{Triana} and smartphone technologies to determine the spring constant using ambient-light sensors,\cite{Pili} magnetic-field sensors,\cite{Pili2019} and smartphone cameras.\cite{Li} These approaches have considerably expanded the range of accessible experiments while reducing the need for dedicated laboratory instrumentation. Nevertheless, most of them remain focused on determining the oscillation period or the spring constant, whereas more comprehensive investigations of the system dynamics often require additional data processing or specialized software.

Among computer-based approaches, video-analysis software such as Tracker\cite{tracker} has become one of the most widely adopted tools for mechanics laboratories because it allows quantitative extraction of motion from recorded videos. However, its use generally involves recording the experiment first and analyzing the video afterward, making immediate interaction with the observed motion less practical. More recently, real-time computer-vision systems based on OpenCV\cite{open-cv} (Open Source Computer Vision) have enabled automatic real-time tracking using ordinary smartphone or webcam cameras. One example is the Pendulum Tracker application,\cite{pendulum-tracker} which performs real-time measurements of pendular motion directly in a web browser.

In this work, we present a low-cost laboratory for investigating the dynamics of a vertical mass--spring oscillator using automatic video tracking. Implemented as a browser-based application within the SimuFísica educational platform, the system automatically tracks the position of the oscillating mass, displays displacement-versus-time graphs during the experiment, and exports the recorded data for further analysis. We first describe the experimental apparatus and the real-time tracking system, and then present a series of representative laboratory activities illustrating the capabilities of the setup. These include damped oscillations, the dependence of the oscillation period on the effective mass, spectral analysis, phase-space trajectories, and nonlinear coupling between spring and pendular modes.

\section{Experimental setup}
\label{experimental-setup}

\subsection{Apparatus}

\begin{figure}[h!]
	\centering
	\includegraphics[width=3.3in]{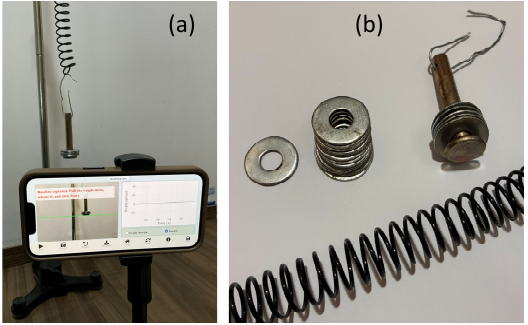}
	\caption{(a) Experimental mass--spring apparatus positioned in front of a smartphone camera. (b) Main components used in the setup: plastic binding coil, steel rod, steel wire, and steel washers.}
	\label{apparatus}
\end{figure}

The experimental apparatus consists of a vertical mass--spring oscillator constructed from inexpensive, commercially available components (Fig.~\ref{apparatus}). The spring is a plastic binding coil measuring 30.2 cm in length, 2.0 cm in diameter, with a mass of 8.5 g and 47 turns. The oscillating mass consists of a steel rod connected to the spring by a short piece of steel wire, with a total mass of 38.8 g. Up to 20 steel washers can be inserted into the rod, allowing the oscillating mass to be adjusted by adding or removing individual washers. Alternatively, the rod-and-washer assembly can be replaced by small padlocks with masses between approximately 30 and 130 g.

The spring is attached directly to an adjustable laboratory support. The total oscillating mass is measured using a digital balance with a resolution of 0.1 g.

A smartphone mounted on a universal holder is used for video acquisition. Except for the smartphone and the laboratory support, all components can be readily purchased from ordinary hardware or stationery stores. Table~\ref{costs} summarizes the approximate cost of the apparatus.

\begin{table}[h!]
	\centering
	\caption{Approximate cost of the experimental components. Prices were estimated from local retail values in Brazil (2026) and converted to U.S. dollars.}
	\begin{ruledtabular}
		\begin{tabular}{lc}
			Component & Approx. cost (US\$) \\
			\hline
			Smartphone holder & 4 \\
			Set of steel washers & 2 \\
			Steel rod & 1 \\
			Steel wire & 0.1 \\
			Plastic binding coil & 1 \\
			Digital balance (0.1 g resolution) & 6 \\
			\hline
			Total (excluding the laboratory support) & $\approx 14$ \\
		\end{tabular}
	\end{ruledtabular}
	\label{costs}
\end{table}

\subsection{Real-time tracking system}

\begin{figure}[h!]
	\centering
	\includegraphics[width=3.3in]{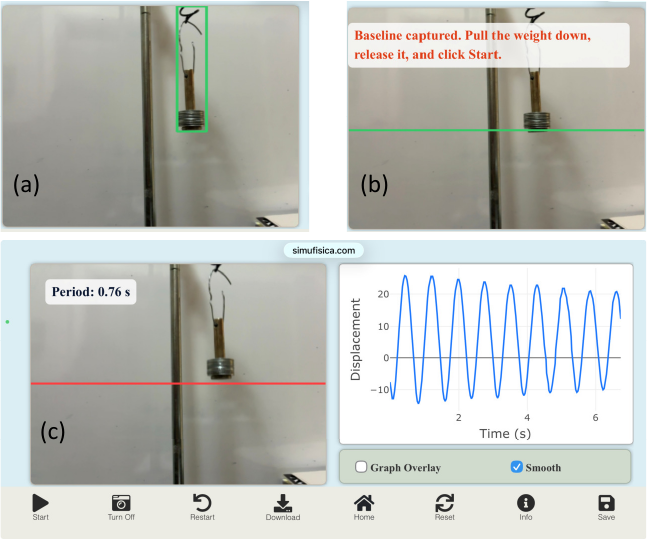}
	\caption{The \textit{Real-Time Mass--Spring Oscillator} application. (a) Automatic detection of the lower portion of the mass--spring assembly after video capture is enabled. (b) Calibration of the equilibrium position, indicated by the horizontal green line below the mass. (c) Real-time display of the displacement-versus-time graph during data acquisition. The application is available at \protect\url{https://simufisica.com/en/real-time-mass-spring-oscillator/}.}
	\label{app}
\end{figure}

Experimental data are acquired using the \textit{Real-Time Mass--Spring Oscillator} application, developed as part of the SimuFísica educational platform (Fig.~\ref{app}). The application uses the OpenCV library to automatically detect the lower portion of the mass--spring assembly through the device camera and track the position of its bottom edge in real time. Its operating principle is similar to that of the SimuFísica \textit{Pendulum Tracker} application~\cite{pendulum-tracker}, which also employs computer vision to identify a moving object and convert its position into a time series. Unlike Pendulum Tracker, however, which tracks approximately the geometric center of the detected object, the present application uses the lower boundary of the mass as the tracking reference, making the measurement largely independent of the geometry of the attached mass.

To perform a measurement, the user first positions the apparatus in front of the device camera and enables video capture. The application then automatically detects the lower portion of the mass--spring assembly [Fig.~\ref{app}(a)]. To calibrate the equilibrium position, the mass is given a slight lateral motion while vertical displacements are kept as small as possible to allow the application to identify the object while keeping the vertical position essentially unchanged. The detected equilibrium position is indicated by a horizontal green line drawn immediately below the mass, allowing the user to verify the calibration visually [Fig.~\ref{app}(b)]. If necessary, the procedure can be repeated using the \texttt{Restart} button.

After calibration, the mass is pulled downward and released while data acquisition is started by pressing the \texttt{Start} button. The application displays, in real time, the displacement relative to the equilibrium position, $\Delta x$, as a function of time [Fig.~\ref{app}(c)]. During the measurement, the reference line changes from green to red and continuously follows the bottom edge of the oscillating mass. Experimental data can be exported using the \texttt{Download} button. The generated text file contains the acquisition time, the measured displacement, and an estimate of the instantaneous velocity computed by finite differences of the displacement signal. These data can subsequently be analyzed using software such as OriginLab, Microsoft Excel, or similar programs.

\section{Example laboratory activities}
\label{representative-experiments}

\subsection{Damped oscillations}
\label{damped-oscillations}

\begin{figure}[h!]
	\centering
	\includegraphics[width=3.3in]{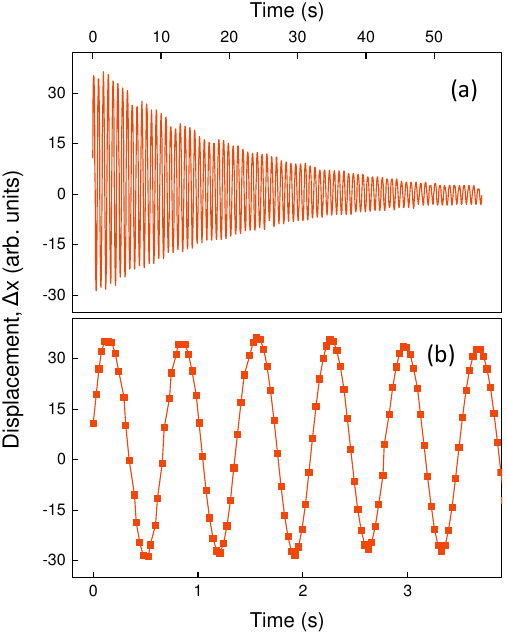}
	\caption{(a) Displacement of the oscillating mass as a function of time. (b) Expanded view of panel (a) over the interval $t\in[0,4]$~s.}
	\label{fig3}
\end{figure}

Figure~\ref{fig3}(a) shows the displacement of the oscillating mass as a function of time for a damped oscillation. The measurement was performed using a total oscillating mass of 62.3 g and a spring extension at equilibrium of $x_0 = 12.5$ cm, measured with a tape measure. The complete record contains 1706 automatically acquired experimental points, illustrating one of the main advantages of the proposed system over conventional video-analysis software such as Tracker, for which manual frame-by-frame digitization of such a large data set would be impractical.

Figure~\ref{fig3}(b) shows a magnified view of the first four seconds of the measurement, corresponding to approximately five oscillation cycles. The graph reveals the high temporal resolution of the acquisition, approximately 28~ms for the smartphone used in this experiment (iPhone 16e). Since the application processes the video stream in real time, the sampling interval depends on the processing capability of the device and may vary with the hardware.

During data acquisition, the application continuously updates the average oscillation period using the five most recent cycles, providing immediate feedback to the user. In the present experiment, the displayed value fluctuated between 0.70 and 0.71~s. This result agrees well with the theoretical period, obtained from

\begin{equation}
	mg = kx_0
\end{equation}

and

\begin{equation}
	T_{\text{theo}} = 2\pi\sqrt{\frac{m}{k}},
	\label{T-mass}
\end{equation}

\noindent where $m$ is the total oscillating mass, $k$ is the spring constant, $x_0$ is the spring extension at equilibrium, and $g=9.78~\mathrm{m\,s^{-2}}$ is the local acceleration due to gravity. Eliminating $k$ using Hooke's law yields

\begin{equation}
	T_{\text{theo}} = 2\pi\sqrt{\frac{x_0}{g}},
\end{equation}

\noindent which gives $T_{\text{theo}}=0.710$~s for the present experiment, corresponding to a relative difference of only 1.4\%.

When Eq. (\ref{T-mass}) is used, the oscillating mass $m$ should include not only the suspended load but also the portion of the spring below the fixed support, which contributes to the system's inertia. Finally, fitting the displacement curve to the damped harmonic oscillator model provides a damping coefficient of $\gamma=0.048~\mathrm{s^{-1}}$, illustrating another experimental activity that can be performed using the recorded data.

\subsection{Oscillation period and the spring effective mass}

A second experiment consists of investigating the dependence of the oscillation period on the effective mass of the system. The effective mass was varied by changing the number of steel washers attached to the oscillating body, while all other experimental conditions were kept unchanged.

\begin{figure}[h!]
	\centering
	\includegraphics[width=3.3in]{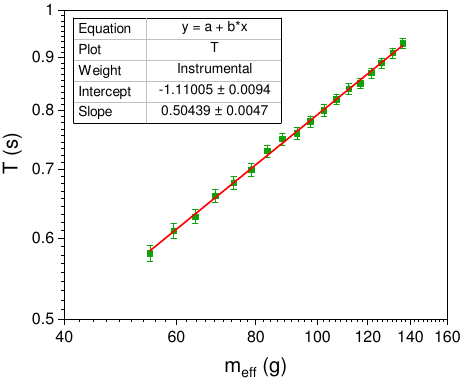}
	\caption{Oscillation period as a function of the effective mass of the system. The red line is a linear regression of the data in the log--log representation.}
	\label{fig4}
\end{figure}

Figure~\ref{fig4} presents the oscillation period as a function of the effective mass on a log--log scale. Here, the effective mass, $m_{\mathrm{eff}}$, is defined as the sum of the suspended mass (steel rod, steel wire, and washers) and the portion of the spring located below the fixed support. For the present apparatus, this contribution corresponds to $8.5\times(34/47)$~g, where 34 and 47 denote the number of active and total turns of the spring, respectively.

The experimental period was obtained directly from the value displayed by the application during the measurement. Although a more precise determination could be obtained from a Fourier analysis of the displacement signal, the adopted procedure was chosen because it is readily accessible in introductory laboratory courses, where students are generally not yet familiar with spectral analysis. The linear regression shown in Fig. \ref{fig4} yields a slope of $0.504 \pm 0.005$, in excellent agreement with the theoretical exponent of $1/2$ predicted by Eq. (\ref{T-mass}).

\subsection{Spectral analysis and phase space}

The dynamics become considerably richer when the oscillation is examined in the frequency domain. While the elementary treatment assumes an ideal Hookean spring, real helical springs exhibit nonlinear restoring forces for sufficiently large deformations \cite{Ivchenko, Ilijić}. As a result, the oscillation is no longer purely sinusoidal, and additional frequency components appear in the Fourier spectrum.

For the present discussion, it is sufficient to retain the leading nonlinear correction to Hooke's law, writing the restoring force as

\begin{equation}
	F(x) = -kx - k_2x^2 + \cdots
\end{equation}

\noindent where the first term represents the familiar linear response and the second term is the leading nonlinear contribution. This quadratic nonlinearity generates new frequency components through wave mixing. In general, these components have frequencies

\begin{equation}
	f_i \pm f_j,
\end{equation}

\noindent where $f_i$ and $f_j$ denote frequencies already present in the spectrum \cite{Boyd}. For an oscillation initially dominated by a single frequency $f$, the most important generated components are $2f$, corresponding to the second harmonic, and $f-f=0$, which produces a zero-frequency (DC) component. Higher-order harmonics may also arise through successive wave-mixing processes (cascading) \cite{Kippenberg}.

\begin{figure}[h!]
	\centering
	\includegraphics[width=3.3in]{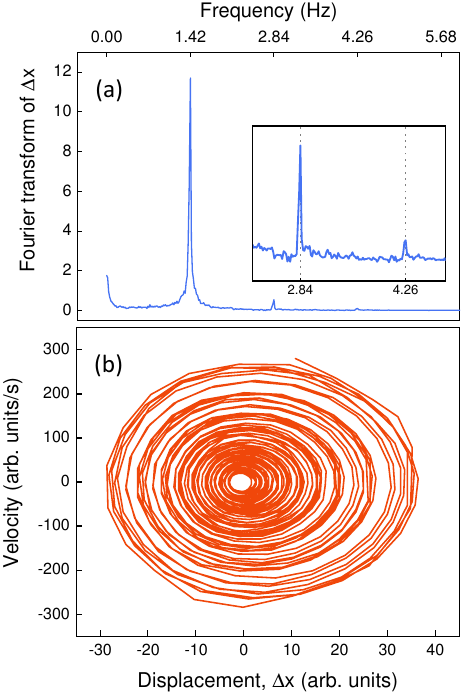}
	\caption{(a) Fourier transform of the displacement signal shown in Fig.~\ref{fig3}(a). Inset: enlarged view of the region $f\in[2.2,4.8]$~Hz. (b) Corresponding phase-space trajectory.}
	\label{fig5}
\end{figure}

Figure~\ref{fig5}(a) shows the Fourier transform of the displacement curve presented in Fig.~\ref{fig3}(a). The spectrum is dominated by the fundamental frequency,

\begin{equation}
	f = \frac{1}{T} = 1.42 \text{ Hz},
\end{equation}

\noindent but also exhibits a pronounced DC component together with clearly resolved peaks at the second ($2f=2.84$ Hz) and third ($3f=4.26$ Hz) harmonics.

The simultaneous observation of these spectral components provides direct experimental evidence of the nonlinear response of the spring. More broadly, it illustrates that a simple mechanical oscillator can exhibit the same wave-mixing mechanisms that underlie many nonlinear optical phenomena \cite{Boyd}.

The corresponding phase-space trajectory is shown in Fig. \ref{fig5}(b). For an ideal harmonic oscillator, the expected trajectory is a symmetric ellipse centered at the equilibrium position. The slight asymmetry observed experimentally is consistent with the nonlinear behavior revealed by the Fourier spectrum, providing an independent visualization of the departure from the ideal Hookean model.

\subsection{Coupling between spring and pendular modes}

Besides behaving as a vertical mass--spring oscillator, the suspended mass can also oscillate as a simple pendulum. Consequently, the system possesses two natural modes of oscillation. As shown in Sec. \ref{damped-oscillations}, the period of the vertical oscillation is

\begin{equation}
	T_s=2\pi\sqrt{\frac{x_0}{g}},
\end{equation}

\noindent where $x_0$ is the spring extension at equilibrium. Since the equilibrium length of the stretched spring, $x$, also corresponds to the pendulum length, the pendular period is

\begin{equation}
	T_p=2\pi\sqrt{\frac{x}{g}}.
\end{equation}

For the present experiment, the equilibrium length was $x = 41.2$ cm and the spring extension was $x_0=9.0$~cm. These values give theoretical periods of $T_p = 1.29$ s and $T_s = 0.603$ s, corresponding to frequencies of $f_p=0.775$~Hz and $f_s=1.659$~Hz, respectively. The system therefore satisfies approximately the condition $f_s\approx2f_p$, for which strong coupling between the vertical and pendular modes is expected.

\begin{figure}[h!]
	\centering
	\includegraphics[width=3.3in]{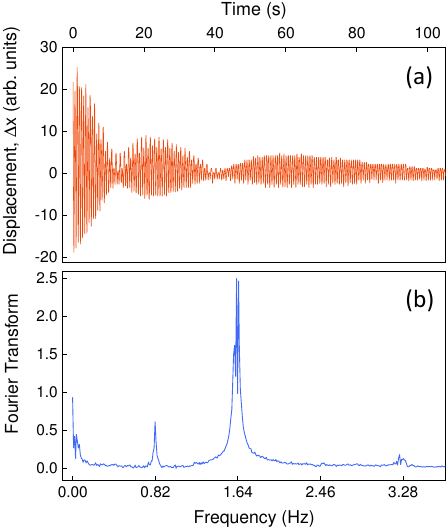}
	\caption{(a) Vertical displacement of the oscillating mass as a function of time. (b) Corresponding Fourier spectrum.}
	\label{fig6}
\end{figure}

The experimental procedure was identical to that used throughout this work. The mass was displaced only in the vertical direction and then released, without any intentional lateral excitation. Nevertheless, after only a few oscillation cycles, a pendular motion developed spontaneously. Figure~\ref{fig6}(a) shows the measured vertical displacement as a function of time. Instead of a simple exponentially damped oscillation, the signal exhibits a slowly varying envelope: the oscillation amplitude decreases almost to zero, increases again, and then decreases once more. This beating pattern provides clear experimental evidence of the periodic exchange of energy between the vertical and pendular modes.

The Fourier spectrum shown in Fig. \ref{fig6}(b) confirms the simultaneous presence of both oscillation modes, with dominant peaks at approximately 0.819~Hz and 1.628~Hz, corresponding to the pendular and vertical oscillations, respectively. A pronounced peak is also observed near 3.246~Hz, corresponding to the second harmonic of the vertical oscillation.

In addition, weaker but reproducible peaks appear near the frequencies expected from nonlinear wave mixing. In particular, a small peak is observed close to the sum frequency $f_s+f_p$, while reproducible low-frequency components are consistent with the combination frequency $2f_p-f_s$, which lies very close to zero. The latter is also consistent with the slowly varying envelope observed in the time-domain signal [Fig.~\ref{fig6}(a)]. Together, these observations are consistent with nonlinear frequency mixing arising from the coupling between the vertical and pendular modes.

\section{Conclusions}

We have presented a low-cost experimental apparatus for investigating the nonlinear dynamics of a vertical mass--spring oscillator. Using computer vision, the system automatically tracks the motion of the oscillating mass and displays its displacement as a function of time while simultaneously providing real-time estimates of the oscillation period. The recorded data can also be exported for further analysis, making the apparatus suitable for both real-time demonstrations and quantitative laboratory experiments.

The representative experiments presented in this work demonstrate that the proposed apparatus supports a broad range of undergraduate laboratory activities, including damped oscillations, measurements of oscillation periods, spectral analysis, phase-space reconstruction, and the investigation of nonlinear coupling between the spring and pendular modes. Owing to its low cost, ease of construction, and automated real-time data acquisition, the proposed apparatus extends the traditional mass--spring experiment to include topics such as Fourier analysis, nonlinear frequency mixing, and coupled oscillations, making these concepts readily accessible in undergraduate laboratories.

\begin{acknowledgments}

This work was supported by the Conselho Nacional de Desenvolvimento Científico e Tecnológico (CNPq, Grant No. 304017/2022-1), the Fundação de Amparo ao Desenvolvimento das Ações Científicas e Tecnológicas e à Pesquisa do Estado de Rondônia (FAPERO, Grant No. 36214.577.20546.20102023), and the Universidade Federal de Rondônia (UNIR, Grant No. 23118.006316/2024-79).
	
\end{acknowledgments}


\begin{thebibliography}{0}%
\makeatletter
\providecommand \@ifxundefined [1]{%
 \@ifx{#1\undefined}
}%
\providecommand \@ifnum [1]{%
 \ifnum #1\expandafter \@firstoftwo
 \else \expandafter \@secondoftwo
 \fi
}%
\providecommand \@ifx [1]{%
 \ifx #1\expandafter \@firstoftwo
 \else \expandafter \@secondoftwo
 \fi
}%
\providecommand \natexlab [1]{#1}%
\providecommand \enquote  [1]{``#1''}%
\providecommand \bibnamefont  [1]{#1}%
\providecommand \bibfnamefont [1]{#1}%
\providecommand \citenamefont [1]{#1}%
\providecommand \href@noop [0]{\@secondoftwo}%
\providecommand \href [0]{\begingroup \@sanitize@url \@href}%
\providecommand \@href[1]{\@@startlink{#1}\@@href}%
\providecommand \@@href[1]{\endgroup#1\@@endlink}%
\providecommand \@sanitize@url [0]{\catcode `\\12\catcode `\$12\catcode
  `\&12\catcode `\#12\catcode `\^12\catcode `\_12\catcode `\%12\relax}%
\providecommand \@@startlink[1]{}%
\providecommand \@@endlink[0]{}%
\providecommand \url  [0]{\begingroup\@sanitize@url \@url }%
\providecommand \@url [1]{\endgroup\@href {#1}{\urlprefix }}%
\providecommand \urlprefix  [0]{URL }%
\providecommand \Eprint [0]{\href }%
\providecommand \doibase [0]{https://doi.org/}%
\providecommand \selectlanguage [0]{\@gobble}%
\providecommand \bibinfo  [0]{\@secondoftwo}%
\providecommand \bibfield  [0]{\@secondoftwo}%
\providecommand \translation [1]{[#1]}%
\providecommand \BibitemOpen [0]{}%
\providecommand \bibitemStop [0]{}%
\providecommand \bibitemNoStop [0]{.\EOS\space}%
\providecommand \EOS [0]{\spacefactor3000\relax}%
\providecommand \BibitemShut  [1]{\csname bibitem#1\endcsname}%
\let\auto@bib@innerbib\@empty
\end{thebibliography}%


\begin{thebibliography}{99}

\bibitem{Triana} C. A. Triana, F. Fajardo, ``Experimental study of simple harmonic motion of a spring-mass system as a function of spring diameter,'' \textit{Rev. Bras. Ensino Fís.} \textbf{35}, 4305 (2013).

\bibitem{Pili} Unofre Pili, ``A dynamic-based measurement of a spring constant with a smartphone light sensor,'' \textit{Physics Education} \textbf{53}, 033002 (2018).

\bibitem{Pili2019} Unofre Pili, Renante Violanda, ``Measuring a spring constant with a smartphone magnetic field sensor,'' \textit{Phys. Teach.} \textbf{57}, 198–199 (2019).

\bibitem{Li} Yujie Li, Jiasheng Wu, Yuhe Zhao, Yu Hu, Jiawei Song, Wei Zhuang, ``A smartphone-based simple method for determination of the spring constant,'' \textit{Phys. Teach.} \textbf{63}, 798–799 (2025).

\bibitem{tracker} D. Brown, W. Christian and R. M. Hanson, ``Tracker: video analysis and modeling tool. (available at: \url{https://opensourcephysics.github.io/tracker-website/}).

\bibitem{open-cv} Intel, Willow Garage 2000 OpenCV: open source computer vision library (available at: \url{https://opencv.org/}.)

\bibitem{pendulum-tracker} Marco P M de Souza, Juciane G Maia and Lilian N de Andrade, ``Pendulum Tracker--SimuFísica: a web-based tool for real-time measurement of oscillatory motion,''
\textit{Phys. Educ.} \textbf{60}, 055024 (2025).

\bibitem{Ivchenko} Vladimir V. Ivchenko, ``Theoretical study of the geometrical non-linearity of the elastic properties of helical springs,'' \textit{Am. J. Phys.} \textbf{88}, 958-961 (2020).

\bibitem{Ilijić} Saša Ilijić, Ana Babić, Dora Ivrlač, Andrew DeBenedictis, ``The nonlinearity of helical springs: An energy-based approach,'' \textit{Am. J. Phys.} \textbf{93}, 932-942 (2025).

\bibitem{Boyd} R. W. Boyd, \textit{Nonlinear Optics}  (Academic Press, 4th ed, 2020).

\bibitem{Kippenberg} T. J. Kippenberg, R. Holzwarth, S. A. Diddams, ``Microresonator-based optical frequency combs,'' \textit{Science} \textbf{332}, 555 (2011).


\end{thebibliography}
\end{document}